\documentclass[11pt]{article}
\usepackage{lineno,hyperref}
\usepackage{placeins}

\usepackage{graphicx,amsthm,amssymb,amsmath,authblk}
\usepackage{hyperref}

\usepackage{algorithm} 
\usepackage[noend]{algorithmic}

\usepackage{amsgen,amsmath,amstext,amsbsy,amsopn,amsfonts,amssymb,stmaryrd,amsthm}
\usepackage{tcolorbox}
\usepackage[english]{babel}
\usepackage{tikz}
\usetikzlibrary{chains,positioning,scopes,shapes} 
\usepackage{color}
\usepackage{caption}
\usepackage{subcaption}
\usepackage{xspace}
\usepackage{mathtools}

\newcommand{\remove}[1]{}

\usepackage{xspace}

\newcommand{\vbw}{{\rm vbw}}
\newcommand{\bw}{{\rm bw}}

\newcommand{\ms}[1]{{\color{magenta}{ \bf [~Maria:\ }\emph{#1}\textbf{~]}}}

\newcommand{\tred}[1]{\textcolor{red}{#1}}
\newcommand{\tblue}[1]{\textcolor{blue}{#1}}

\title{\scshape On Vertex Bisection Width of Random $d$-Regular Graphs}

\author[1,4]{Josep D\'{\i}az\thanks{Research supported by the Spanish Agencia Estatal de Investigación under project PID2020-112581GB-C21}}
\author[2,3]{\"{O}znur Ya\c{s}ar Diner\thanks{ \"O. Y. Diner is partially supported by  the Scientific and Technological Research Council T\"ubitak under project BIDEB 2219-1059B191802095 and by Kadir Has University under project 2018-BAP-08}}
\author[1,4]{{Maria Serna$^{*}$}}
\author[1,4]{Oriol Serra\thanks{Research supported by the Spanish {\it Agencia Estatal de Investigaci\'on} under project PID2020-113082GB-I00.}}
\affil[1]{ALBCOM Research Group, Computer Science Department.  Universitat Polit\`{e}cnica de Catalunya, Barcelona, Spain}
\affil[2]{Computer Engineering Department, Kadir Has University, Istanbul, Turkey}
\affil[3]{Mathematics Department.  Universitat Polit\`{e}cnica de Catalunya, Barcelona, Spain}
\affil[4]{Institut de Matem\`{a}tiques de la UPC-BarcelonaTech (IMTech), Universitat Polit\`{e}cnica de Catalunya,  Barcelona}

\begin{document}

\maketitle

\begin{abstract}
Vertex bisection is a graph partitioning problem in which the aim is to find a partition into two equal parts that minimizes the number of vertices in one partition  set that have a neighbor in the other set. We are interested in giving upper bounds on the vertex bisection width of random $d$-regular graphs for constant  values of $d$. Our approach is based on analyzing a greedy algorithm  by using the Differential Equations Method. In this way, we obtain the first known upper bounds for the vertex bisection width in random regular graphs. The results are compared with experimental ones and with lower bounds obtained by Kolesnik and Wormald, (Lower Bounds for the Isoperimetric Numbers of Random Regular Graphs,
SIAM J. on Disc. Math. 28(1), 553-575, 2014).
\end{abstract}


\smallskip

\section{Introduction}
Partitioning the vertices of a given graph $G$ into subsets, subject to certain constraints, generates an important collection of  algorithmic problems, which are used to model a large  range of practical  applications, see e.g.~\cite{Bichot2011}.  

The quest for designing fast algorithms for partition models has generated an important body of theoretical   study of graph partitioning problems. 

One of the partition problems that has  received considerable attention, from both the theoretical and the experimental point of view,  is the {\em Minimum Edge Bisection} as, among other applications,  it measures the bottleneck on the  amount of information that can flow through a network.  

Let $G = (V, E)$ be a graph with $n$ vertices. For any $S\subseteq V$, with $|S|=\lfloor n/2\rfloor$, the pair $(S, V\setminus S)$ is a partition of the vertex set $V$ into equal sized sets when $n$ is even and differing by one when $n$ is odd.  Such a partition is called a \emph{bisection} of the graph. The \emph{edge width} $\bw(S,V\setminus S)$ of a bisection $(S, V\setminus S)$  is the number of edges connecting vertices in its different  parts. The minimum edge bisection or {\em bisection width} of $G$ is the minimum width over all bisections:
$$\bw (G)= \min\{\bw((S, V\setminus S)): (S, V\setminus S) \text{ is a bisection}\}.$$

The corresponding decision problem, {\sc Minimum edge bisection}, has been a very productive research area, using  theoretical and empirical methods. 

Concerning its algorithmic complexity, {\sc Minimum edge bisection} is NP-complete  for general graphs \cite{Garey1976} and it remains in NP-complete when restricted to many graph classes; in particular $d$-regular graphs even for $d=3$~\cite{Bui1987},  grids with an arbitrary number of holes and their equivalent class of unit disk graphs~\cite{Diaz2017}. The problem is efficiently solvable for trees \cite{Nakano2003}, hypercubes \cite{Mac1978},  grid graphs with a constant 
number of holes~\cite{Feldmann11} and  on graphs with bounded treewidth~\cite{Jansen2005}. 
Regarding poly-time approximation algorithms,  it is known that there exists no polynomial time approximation scheme to approximate bisection width within a constant factor~\cite{Khot2004}.   The best approximation algorithm for the problem, achieves a ratio of $O(log n)$ \cite{Racke2008}. From the practical point of view, there has been a slow improvement of efficient heuristics, from the simulated annealing  \cite{Kerningan70} to  one that in practice is very efficient~\cite{Delling2012}. 
There are also various software packages dedicated to solving partitioning problems, including the {\sc Minimum edge bisection}, see for example   METIS~0.7.3 \cite{Karypis1998}. 

A fruitful line of research has been the study of  asymptotic tight bounds for the bisection width of random $d$-regular graphs, in particular for cubic graphs ($d=3$). Buser~\cite{Buser1984}  showed the existence of a cubic graph  $G$ with $\bw(G)>\frac{n}{256}$. Bollob\'as~\cite{Bollobas1988} extended Buser's  result to show that for almost all cubic graphs $G$ one has  $\bw(G)>\frac{n}{11}$, and obtained a lower bound for $d$-regular graphs for all $d\ge 3$. Clark and Entringer~\cite{Clarke1989}  showed that for almost all cubic graphs, $\bw(G)\le \frac{(n+138)}{3}$. They also proved  that almost every $d$-regular $G$  $c_dn \leq \bw(G)$ where $c_d\to d/4$  as $d\to \infty$, which is asymptotically 
correct. In the last 30 years, several  improvements have tightened the bounds for bisection width. Alon \cite{Alon1997} give upper bounds for the edge bisection width for general values of the degree $d$. The method in \cite{Alon1997} can be turned into a randomized algorithm to obtain bisections with small edge width. These bounds were later improved by Lyons \cite{Lyons2017}. Lichev and Mitsche \cite{Lichev2020} obtained very tight upper and lower bounds for the cubic case. D\'iaz, Serna and Wormald \cite{Diaz2007} gave a randomized greedy algorithm for constructing bisections with small edge width  which provide, for $4\le d\le 12$, the best  known asymptotically almost surely (a.a.s.) upper bounds for the edge-bisection of random $d$-regular graphs.   Recall that an event is said to happen a.a.s. if it happens with probability tending to 1 when $n\to \infty$. In \cite{Diaz2007}, the greedy algorithm is analyzed by means of the Differential Equation Method (DEM). The DEM, introduced by Kurtz \cite{Kurtz1970},  has been used  in various combinatorial optimization problems~\cite{Karp1981,Diaz2010}. 
The general framework to use the DEM to bound the solution to a given problem is to design a greedy algorithm that at each step increments the size of a solution; if $X_t$ is a random variable indicating the size of the solution at step $t$,  we consider   the expected change of $|X_{t+1}-X_t|$ in one step of the algorithm,  and define the differential equation indicated by this expected change. The solution to that differential equation is the expected size of the solution. Moreover, as the stochastic sequence of those random variables should form a semi-martingale, the expected value is concentrated, giving an approximate solution to the problem, see e.g.  Wormald~\cite[Theorem 5.1]{Wormald1999b}.

In this paper we adapt this  strategy to obtain upper bounds on the  {\em vertex bisection width} of random $d$-regular graphs for $d$ constant. 

Given  a graph $G=(V,E)$ and a partition $(S, V\setminus S)$, let $S^*=\{u\in S: \exists v\in V\setminus S, uv\in E\}$ and let
let $\overline{S}^*=\{v\in V\setminus S : \exists w\in S,~{\rm with~} wv\in E\}$.\\
The \emph{vertex width} of  the bisection $(S, V\setminus S)$ is defined as 
$$\vbw((S, V\setminus S))=\min \{|S^{*}|, |\overline{S}^*|\},$$ 
and the \emph{vertex bisection width} of $G$ is defined as 
$$
\vbw(G)= \min_{\{(S, V\setminus S)\}}\{\vbw((S, V\setminus S))\}.
$$
In the {\sc Min vertex bisection problem} the aim is to find the bisection  with minimal bisection width.

We note that  a slightly different definition of {\em vertex bisection width}, which  considers instead the  minimum number of vertices to remove in order to obtain a disconnected partition of the graph in two equal sized  subgraphs, can be found in the literature, see for example~\cite{vanBevern2015}. In this paper we will stick to the former, somewhat more natural, definition of vertex bisection width. 

It is straightforward to obtain  the following trivial bounds connecting edge and vertex bisection widths for  any $d$-regular graph $G$:
$$
\bw(G)/d\le \vbw(G)\le \bw(G).
$$
The problem of estimating the vertex bisection width of $d$-regular graphs asks for the appropriate value of $c\in [1/d,1]$ such that $\vbw(G)=c\cdot \bw(G)$. 

Both the edge and vertex bisection problems are particular instances of the edge and vertex graph isoperimetric problems, where the minimization is taken over all sets of a given cardinality (not necessarily $n/2$). The example of the $n$-cube, where an exact solution for the two problems is known (see Harper~\cite{Harper1964,Harper1966}) illustrates the different structural behaviour of the two problems. Roughly speaking, the solutions for the edge isoperimetric problem are close to subspaces of the space of binary sequences, which turn out to  maximize vertex boundary, whereas the ones for the vertex isoperimetric problem are close to balls, which maximize edge boundary.  

In spite that the  {\sc Min Vertex Bisection} arises in important applications, as  the study of message passing and fault tolerance in interconnection networks, see e.g.~\cite{Diaz2002,Sauerwald2011}, the problem has received little attention, in particular from the theoretical point of view.  

The problem is known to be NP-complete for general graphs and it is solvable in polynomial time for hypercubes~\cite{Brandes2009}. There have been heuristics based in integer linear programming  and branch and bound~\cite{Fraire2014,Jain2016}, where due to the exhaustive search nature of the techniques, their applicability is reduced only to graphs of small size. 

Recently, Kolesnik and Wormald \cite{Kolesnik2014} obtained a.a.s.  lower bounds for the vertex bisection width of random $d$-regular graphs for every $d\ge 3$. 

One of the main motivations of this paper is to provide corresponding upper bounds by producing bisections with small width. This is achieved by proposing a  randomized greedy algorithm which is suited for  the vertex bisection problem, and analyze it by means of the Differential Equation Method.   

The paper is organized as follows. In Section \ref{sec:greedy} we describe a greedy algorithm to obtain a partition with small vertex bisection of a $d$-regular graph,  which provides upper bounds for the parameter. In section \ref{sec:dem}, we present a modification of the previous algorithm, suitable for its analysis by the Differential Equations Method, when applied to random $d$-regular graphs.  The results and discussion, both experimental and by the differential equation method, are presented in Section \ref{sec:results}. The paper concludes in section \ref{sec:conclusions} with some final remarks and lines of future research. In the Appendix we provide additional information on the experimental part of our contribution.    

\section{A greedy algorithm}\label{sec:greedy}

Upper bounds for the vertex bisection width  $\vbw(G)$ of a graph $G=(V,E)$  are obtained by exhibiting vertex bisections, or equivalently, by providing balanced Red/Green vertex colorings  $(R,G)$, with $|\, |R|-|G|\, |\leq 1$, with a set $R_0\subset R$ of interior vertices of color red  as large as possible, as the vertex bisection width of such a partition is $|R|-|R_0|$.

For a vertex $x_0$ and an integer $r$,  the ball $B(x_0,r)$ with center $x_0$ and radius $r$  is set of vertices of $G$  at  distance at most $r$ from $x_0$.  A natural guess for bisections with small vertex width  may consider that the set  in the partition will contain a large ball. A second consideration is that, in such a partition, one expects to have vertices with a large number of neighbors on the other side of the partition. These two considerations suggest the greedy Algorithm~\ref{alg:1} to produce suitable vertex bisections with small vertex bisection.

We next explain Algorithm~\ref{alg:1}.  
Let $r_c(x_0)$ be the smallest $r$, $1\leq r\leq n$, such that $|B(x_0,r)|> n/2$. Our algorithm for computing a vertex bisection starts  by coloring RED a ball around a randomly chosen vertex $x_0$ of radius $r_0(x_0)=r_c(x_0)-2$. For simplicity we write $r_0$ and $r_c$ when $x_0$ is clear form the context. We next proceed to enlarge the set $R$ of red vertices  up to 
$n/2$ in a greedy fashion. This is done  by first selecting a colored vertex with the smallest possible number of uncolored neighbors and then by coloring red one of its uncolored neighbors.  The set $R$ of red vertices returned by the algorithm contains a ball centered at $x_0$ of radius $r_0-1$, ensuring a  large subset of interior vertices, so that  the expected bisection width will be  small. In the second phase of  Algorithm~\ref{alg:1}, we color RED new vertices until we get a bisection. In this phase, we greedily select the vertices with the fewer uncolored neighbors.

\begin{algorithm}[t]
 \caption{Greedy algorithm to create a vertex bisection of a given $d$-regular graph $G=(V,E)$} \label{alg:1}
 \begin{algorithmic}
\STATE Input: A $d$-regular graph $G$ with $|V|=n$
\STATE Choose a random vertex $x_0$
\STATE $R=\emptyset$, $B=\{x_0\}$, $D=N(x_0)$ 
\STATE \COMMENT{$R\cup B =B(x_0,0)$, $R\cup B \cup D =B(x_0,1)$ } 
\STATE \COMMENT{Phase 1}
\WHILE{$|R\cup B \cup D| \leq  n/2$} 
\STATE $R = R\cup B$, $B = D$
\STATE $D = \{N(u)\cap (V\setminus(R\cup B)) \mid u\in R\cup B\}$
\ENDWHILE 
\STATE \COMMENT{$R=B(x_0,r_0)$, $R\cup B =B(x_0,r_0+1)$, $R\cup B \cup D =B(x_0,r_c)$} 
\STATE \COMMENT{Phase 2}
\WHILE{$|R|<n/2$} 
\STATE For $0< i <d$, let $R_i=\{u\in R \mid |N(u)\setminus R|=i\}$
\STATE Let $j= \min \{i\mid |R_i|>0\}$
\STATE Choose a random vertex $v$ from $R_j$
\STATE Choose a random vertex $w$ from $N(v)\setminus R$
\STATE $R = R\cup \{w\}$
\ENDWHILE 
\STATE Return $(R,V\setminus R)$.
\end{algorithmic}
\end{algorithm}

We ran versions of the algorithm with distinct choices of $r_0$ to start the second phase of the algorithm. With the choice $r_0=r_c-1$ an undesired irregular behaviour is observed for distinct values of the number $n$ of vertices in random $d$--regular graphs, depending on how close is the  ball $B(x_0,r_c)$ to a set of cardinality $n/2$. The closer this size to $n/2$, the shorter is the range of execution of  the second phase of the algorithm, resulting in larger bisection widths. On the other hand, smaller choices of $r_0$ below $r_c-2$ slow down the execution time with not producing better results. In Table~\ref{tab:d4}, we show the proportion of vertices in the bisection width obtained in a run of Algorithm~\ref{alg:1} with two selections for $r_0$ on graphs of different sizes. Empirically, it can be seen that the choice $r_0=r_c-2$  yields better bisection widths.   In Appendix~\ref{ape:A} we provide several plots comparing the sizes of the balls before and after reaching size $n/2$ as a function of $n$. From the plots, it can be seen that,  as expected, the size of the balls  follow closely a step function. 

\begin{table}
\begin{center}
\begin{tabular}{|c|cccccc|}
\hline
$n$  &$10^5$ & $2\cdot 10^5$ & $3\cdot 10^5$ & $4\cdot 10^5$ & $5\cdot 10^5$& $6\cdot 10^5$\\\hline
$r_0=r_c -2$ &$0.46180$ & $0.46630$ & $0.46135$ & $0.45910 $ &  $0.45905$ &   $0.46594$\\\hline
$r_0= r_c-1$ &$0.47974$ & $0.514479$ & $0.47929$ & $0.473765  $ &  $0.46867$ &   $0.51469$\\\hline
\end{tabular}
\end{center}
\caption{Experimental results of Algorithm~\ref{alg:1},for $d=4$ and different values of $n$, with different selections of $r_0$.}\label{tab:d4}
\end{table} 

Given the graph $G$ as an adjacency list, Algorithm~\ref{alg:1} visits the list of neighbours of only red vertices  at most $d$ times each, so the overall running time of the algorithm is $O(dn)$.   


In our experiments, we use Algorithm \ref{alg:1} (coded in Phyton-3)  to obtain upper bounds for the bisection width of random $d$-regular graphs. Experiments are run on several instances of random regular graphs with number of vertices ranging from  $n=10^5$ to $n=6\cdot 10^5$. We selected this range because it covers, for most of the values of $d$, the highest variability on ball's sizes (see Appendix~\ref{ape:A}). We have used   the generator of random $d$--regular graphs  included in the Python package NetworkX-2.5. The results, reporting the average of $5$ runs of the algorithm on each graph together with the maximum and minimum value are given in Appendix~\ref{ape:B}.  Table~\ref{fig:results} 
in Section~\ref{sec:results}  reports the maximum obtained vertex bisection value for $4\leq d\leq 10$.

\begin{algorithm}[t]
 \caption{Create a description of $B(x_0,r_0+1)$ on a random $d$-regular graph $G$ on $n$ vertices in the pairing model}\label{alg:2}
  \begin{algorithmic}
\STATE Input: $d$ and $V$ with $|V| = n$
\STATE \COMMENT{Notice the random $d$-regular graph is generated as we expose edges}
\FOR{$i=0,\dots, d-1$}
\STATE $R_i = \emptyset$, $Z_i= \emptyset$
\ENDFOR
\STATE Choose a random vertex $x_0$
\STATE $R_0= \{x_0\}$,  $Z_{d-1}= \{N(x_0)\}$, $Z_d= V\setminus (R_0 \cup Z_{d-1})$, $R= R_0$ 
\WHILE{$|R|<n/2$}
\FOR {$i=1$ \TO $d-1$} 
\STATE $R_i=Z_i$, $R{\rm aux}_i =R_i$ 
\STATE{$Z_i=\emptyset$}
\ENDFOR
\STATE $R=\cup_{i=0}^{d-1}R_i$
\STATE $Z{\rm aux}_{d}=Z_{d} $, $R{\rm aux}=R$
\STATE \COMMENT{Phase}
\WHILE{$|R|-|R_0|>0$}  
\STATE Choose a random unpaired point $x$ in vertices from $R\setminus R_0$ 
\STATE Compute the $i$ such that $x\in R_i$  
\STATE Remove $x$ from $R_i$ and add it to $R_{i-1}$
\STATE Choose a random unpaired point $y$ from $V\setminus (R_0\cup Z_0)$
\IF{$y\in R$}
\STATE Compute the value $j$ such that $y\in R_j$ 
\STATE Remove $y$ from $R_j$ and add it to $R_{j-1}$
\ELSE 
\STATE Compute the value $j$ such that $y\in Z_j$ 
\STATE Remove $y$ from $Z_j$ and add it to $Z_{j-1}$
\ENDIF
\STATE \COMMENT{The edge $xy$ has been exposed, and the corresponding sets updated}
\ENDWHILE
\ENDWHILE 
\STATE Return $(G,R{\rm aux},R{\rm aux}_0, \dots , R{\rm aux}_{d-1}, Z{\rm aux}_d)$.
\end{algorithmic}
\end{algorithm}

\section{Analyzing a greedy algorithm with the differential equation method}\label{sec:dem}
\remove{The {\em Differential Equation Method}  is a set of techniques   used  in combinatorics and algorithmic to analyze dynamic random processes that evolve in tiny steps (with respect to the final structure). For example,  one step can be exposing a new edge in the construction of a large graph.  The method produces tight bounds on a  set of random variables that almost asymptotically surely hold at every step. The very useful characteristic of the method is that, at each step, each random variable is concentrated around its expectation as it evolves from step to step, see  Wormald~\cite{Wormald1999} and D\'iaz and Mitsche~\cite{Diaz2010} for a thorough discussion of the method.}

Algorithm~\ref{alg:1} is not   suited for its analysis with the differential equation method  as in the first phase,  at each step all the vertices in the boundary of a ball are colored RED instead of being colored one by one. This makes  the algorithm~\ref{alg:1} more efficient in terms of running time, however the difference of increment   of vertices between two consecutive steps is too large to  transform it into a continuous smooth function, as needed in the DEM. 
 
A more suitable  algorithm  for the application of the differential equation method to the first phase of Algorithm~\ref{alg:1} to be run on  random $d$--regular graphs  is the one presented  in Algorithm~\ref{alg:2}.

We use the pairing model to generate random $d$-regular graphs with $n$ vertices, where $nd$ is assumed to be even. The model was introduced by Bollob\'as~\cite{bollobas1980} and by Wormald~\cite{Wormald1999b}, see Gao and Wormald~\cite{Gao2017} for a very efficient way to generate a random $d$-regular graph. The basic working of the  pairing model chooses a random matching on a set of $nd$ points that are grouped into bags of size $d$, each bag corresponding to a vertex. The resulting object is  a $d$-regular multigraph (it can have loops and multiple edges), but a basic result on the pairing model is that, with positive probability bounded away from zero, one gets a $d$-regular simple graph with the uniform distribution. Therefore it follows that every property holding a.a.s. in the pairing model also holds a.a.s. for random $d$-regular graphs. 

The pairing model can be equivalently described as a graph process $(G_t, t=0,1,\ldots ,nd)$, by arbitrarily choosing at each time $t$  an unpaired point and by pairing it with a randomly chosen unpaired point, {\em exposing} in this way an edge of the graph, at time $t$ (see e.g.~\cite{Wormald1999b}.)

\subsection{Algorithm 2} In this subsection, we first introduce Algorithm~\ref{alg:2}, which is an adaptation of the first phase of Algorithm~\ref{alg:1} to deal with  the pairing model.  The algorithm implements the first phase of  Algorithm~\ref{alg:1} coupled with the  graph process generating a $d$-regular graph in the pairing model. Algorithm~\ref{alg:2} obtains the next ball, by exposing (one by one) the edges out of the current ball. In this way, one step of the algorithm makes only small changes on the status of the RED vertices. In order to deal directly with edges we need to keep additional information about the partially exposed  graph.     

Algorithm~\ref{alg:2}  keeps a partition of $V$  with respect  the the current set of RED vertices:
\begin{itemize}
\item $R_i$ is the  set of red vertices with $i$ unpaired points, $0\le i<d$.
\item $Z_i$ is the set of non red vertices with $i$ unpaired points, $0\le i\le d$.
\item $R = \cup_{0}^{d-1} R_i$, is the set of red vertices.
\end{itemize}

\FloatBarrier 

Algorithm~\ref{alg:2} enlarges a ball of red vertices around $x_0$ by, at each iteration of the inner while,  
choosing randomly one unpaired point from a red vertex, then by choosing randomly an unpaired point, from a red or a non red vertex, and by pairing the two selected points. This operation exposes a new edge. After this, it updates the sets $R$, $R_i$ and $Z_i$.  This basic operation is repeated  as long as (i)  there are unpaired points in RED vertices, and (ii) there are less than $n/2$ RED points.  When Algorithm~\ref{alg:2} runs out of unpaired points but there are less than $n/2$ red points,  we color RED all uncolored points in $\cup_{i=1}^{d-1}Z_i$, updating the variables $R$, $R_i$ and $Z_i$ accordingly.  
Observe that this process is equivalent to grow $B(x_0,r)$ from $B(x_0,r-1)$. A phase of Algorithm~\ref{alg:2}, corresponds to the growth of one additional layer in the ball.

In order to apply the Differential Equation Method to  estimate  the sizes of the different sets in the output of  Algorithm~\ref{alg:2}, we define several  random variables.

\begin{itemize}
\item $r_i=|R_i|$, $0\leq i < d$, counts the number of  red vertices  with $i$ unmatched points.
\item $nR=\sum_{i=0}^{d-1} r_{i}$, counts the number of red vertices. 
\item $pR=\sum_{i=0}^{d-1}  i r_{i} $, counts the number of  unmatched points in vertices colored red.
\item $z_{i}$, $1\le i\le d$, counts the number of non red vertices with $i$ unmatched points.
\item $nZ=\sum_{i=0}^d z_{i}$ counts the number of non red vertices.
\item $pZ=\sum_{i=0}^d iz_{i}$ counts the number of unmatched points in vertices not colored red.
\item $pW=pR+pW$, counts the total number of unmatched points. 
\end{itemize}
All the variables depend on the step $t$ of the algorithm, but this dependence is not made explicit in the notation.

We run Algorithm \ref{alg:2} in phases, each phase being finished when $|R|-|R_0|=0$, which correspond to the inner while loop. The input of the next phase, as long as $|R|<n/2$, is initialized by updating the variables $R_i$ to $Z_i$  and $Z_i$ to zero, $0\le i\le d-1$.

At each step $t$ of the inner loop of the Algorithm~\ref{alg:2}  we discover one edge of the graph. We compute the  expected changes in the values of the variables (up to $O(1/nW)$ terms). For the variables $r_i$, $0\le i<d$, these changes are displayed in the following Table,

\begin{center}
\begin{tabular}{r|ccc}
first point/second point&  $ir_i$ & $(i+1)r_{i+1}$ & $nW-ir_i-(i+1)r_{i+1}$\\\hline
$ir_i$ & $-2$ & $0$ & $-1$\\
$(i+1)r_{i+1}$ & $0$ & $2$ & $1$\\
$pR-ir_i-(i+1)r_{i+1}$ & $-1$ & $1$ & $0$
\end{tabular}
\end{center}
monitoring the two random choices for the points to be paired, the first one among points counted by $pR$ and the second one among all $nW$ available points. The variable $r_i$ is modified if the points are chosen in $R_i$ (decreasing its value) or $R_{i+1}$ (increasing its value). According to the displayed values, the expected change of the variables $r_i$ is
 
\begin{align*}
\Delta (r_i)&=-2\frac{ir_i}{pR}\frac{ir_i}{nW}\\&-\frac{ir_i}{pR}\frac{nW-ir_i-(i+1)r_{i+1})}{nW}-\frac{pR-ir_i-(i+1)r_{i+1}}{pR}\frac{ir_i}{nW}\\
&+\frac{(i+1)r_{i+1}}{pR}\frac{nW-ir_i-(i+1)r_{i+1}}{nW}+\frac{pR-ir_i-(i+1)r_{i+1}}{pR}\frac{(i+1)r_{i+1}}{nW}\\&+2\frac{(i+1)r_{i+1}}{pR}\frac{(i+1)r_{i+1}}{nW},
\end{align*}

A similar analysis applies to the expected change of the variables $Z_i$ for $0\le i<d$,

$$
\Delta (z_i)=-\frac{iz_i}{nW}+\frac{(i+1)z_{i+1}}{nW}.
$$ 
and
$$
\Delta (z_d)=-\frac{dz_d}{nW}.
$$
By applying the differential equation method, these expected changes are translated to the  system of differential equations: 

\begin{align*}
    r'_i&=\frac{1}{nW\cdot pR}(2(ir_i)((i+1)r_{i+1})+(nW+pR)((i+1)r_{i+1}-ir_i)),&0\le i<d,\\
    z'_i&=\frac{1}{nW}((i+1)z_{i+1}-iz_i,&0\le i<d,\\
    z'_d&=-\frac{1}{nW}dz_d. &
\end{align*}

The solution to this system of differential equations at the stopping time  provides an asymptotically good approximation of the values of the variables  on the output of Algorithm \ref{alg:2}, as ensured by  the main result in Wormald \cite[Theorem 5.1]{Wormald1999}.  The applicability of that theorem  requires only that the random variables in the process have bounded variation (in our case these variations are bounded by two in absolute value) and that the corresponding continuous versions are Lipschitz (in our case they are quadratic functions of the variables) together with the natural integrability conditions of the system in the domain of interest.

Several phases of the algorithm governed by the same system of differential equations are run where each phase stops when there are no unmatched red points left. These results are feeding the initial values of the variables in the next phase, until the final phase, which stops when $nR=n/2$. 

To analyze the successive systems of differential equations we need to determine suitable initial conditions. For doing so, we take into account that a   
random $d$-regular graph with $n$ vertices contains a.a.s. a rooted complete $d$-ary tree of depth $k=\log\sqrt{n}$, i.e  all vertices except the root and the leaves have $(d-1)$ children (see Bollob\'as and de la Vega~\cite{Bollobas1982}). More explicitly, Makover and McGowan~\cite{makover2010} showed that a.a.s. a random $d$-regular graph contains a complete $d$-ary tree with at least  $n^{1/2-\epsilon}$ vertices for every $\epsilon>0$.   
The interior vertices of the tree have all its $d$ points already paired. Each leaf has $d-1$ unpaired points. 

For the implementation, we integrate the system of differential equations with the Runge-Kutta solver DOP853 in Python-3 with  $10^6$  steps. The system is initialized with  $\epsilon+\epsilon/(d-1)$ RED vertices, corresponding to a complete $d$-ary tree, with $\epsilon=10^{-5}$ leaves. Thus the initial values of the variables are $r_0=\epsilon/(d-1)$, $r_{d-1}=\epsilon$, $z_d=1-\epsilon-\epsilon/(d-1)$ and the remaining variables are set to zero.

\begin{figure}
\scalebox{0.85}{\vbox{
\begin{center}
\includegraphics[width=5cm]{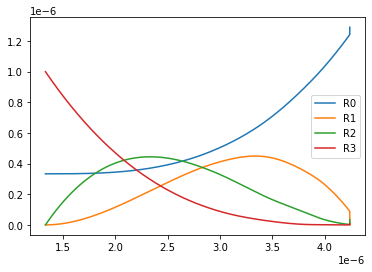}\hspace{5mm}\includegraphics[width=5cm]{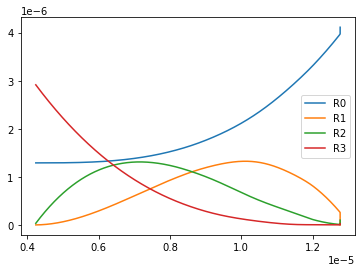}

\includegraphics[width=5cm]{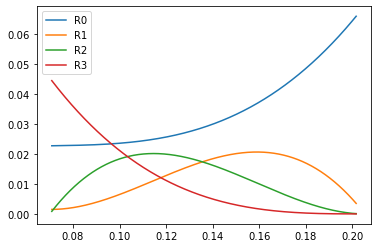}\hspace{5mm}\includegraphics[width=5cm]{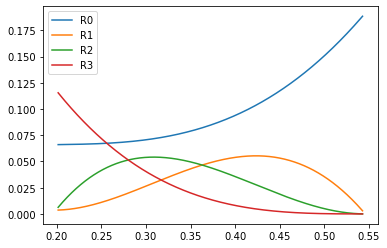}

\includegraphics[width=5cm]{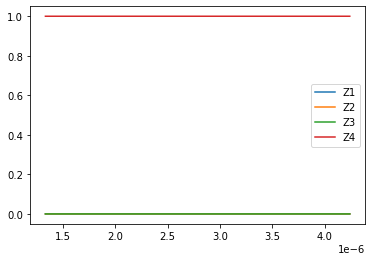}\hspace{5mm}\includegraphics[width=5cm]{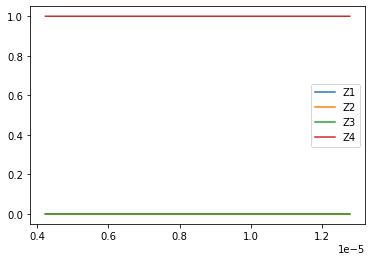}

\includegraphics[width=5cm]{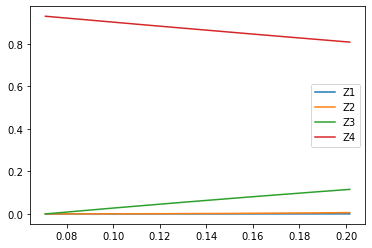}\hspace{5mm}\includegraphics[width=5cm]{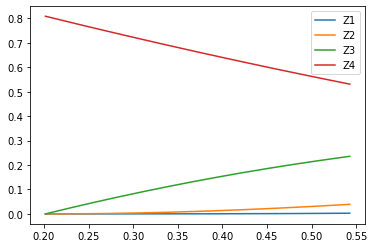}
\end{center}}}

\scalebox{0.85}{\vbox{
\begin{center}
\includegraphics[width=5cm]{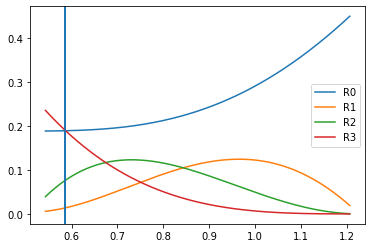} \includegraphics[width=5cm]{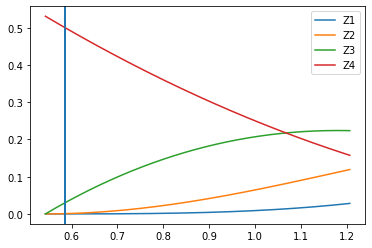}
\end{center}}}
\caption{The evolution of $r_i$ and $z_i$ variables in the solutions to the system of differential equations. The first line contains the first two phases of  the solution of the DEM for Algorithm~\ref{alg:2}, the second line the two last ones. The last two the phase in which the red points form a bisection. \label{fig:plots}}
\end{figure}   

When approximately solving the system of differential equations for a phase in Algorithm~\ref{alg:2}, we do not expect to get exact zero values on all variables. We stop a phase just before one of the variables in the solution gets below zero. Then, the initial values for the following phase are obtained by, for $0\leq i\leq d-1$, accumulating in  $r_i$ the value $z_i$ and then setting to zero $z_i$. We stop the process just before over-passing the 0.5 bound. 
We illustrate, for $d= 4$, the evolution of the $r_i$ and the $z_i$ variables along phases in Figure \ref{fig:plots}. 

\begin{algorithm}[t]
 \caption{Second phase to get a vertex bisection on a random\\ {$d$-regular} graph $G$ on $n$ vertices}\label{alg:3}
\begin{algorithmic} 
\STATE{Input: $(G,R,R_0, \dots R_{d-1}, Z_d)$, a partially exposed graph and the output of Algorithm~\ref{alg:2}}
\STATE $R_d = Z_d$
\STATE $R=\cup_{i=0}^{d-1}R_i$
\STATE $L=\cup_{i=1}^{\lceil d/2 \rceil}R_i$
\WHILE{$|R|-|R_1|<n/2$}
\STATE {Choose a random unpaired point $x$ in vertices from $L$}
\STATE {Choose a random unpaired point $y$  in vertices from $V\setminus R_0$}
\STATE \COMMENT{The Edge $xy$ has been exposed and the corresponding sets updated}
\ENDWHILE 
\STATE Add to $R$ enough neighbors of vertices in $R_1$ to get $|R| = n/2$
\STATE Return $(R\setminus R_1,V\setminus (R\setminus R_1))$.
\end{algorithmic} 
\end{algorithm}
   
\subsection{Algorithm 3} The output of Algorithm~\ref{alg:2} provides the values of the variables in the one before the last round, when we reach $|R|=n/2$. To complete the algorithm we want next  to mimic the second phase of Algorithm~\ref{alg:1} and analyze it with the DEM.

We have been unable to analyze the second phase of Algorithm~\ref{alg:1} via the DEM. Instead, we analyze a variant of the second phase described as Algorithm~\ref{alg:3}. In this variant instead of keeping a strict priority by number of uncovered points, we select the first uncovered point from vertices with $d/2$ or less uncovered points and the second one from all available points. As a minor modification, the stopping time is set to $|R|-|R_1|=n/2$ instead of $|R|=n/2$. This is so because the vertices in $R_1$, when placed in the complement, contribute in at most $|R_1|$ to the vertex bisection width, which provides a slight improvement in the seeked upper bound.

We keep the same notation as in Algorithm~\ref{alg:2} for $R_i$ and $r_i$, $0\le i<d$ but now we denote by
$$
pL=\sum_{i=1}^{\lceil d/2 \rceil}  i r_{i},
$$
the number of  unmatched points in vertices colored red that are used as first point,  by $R_d$ the set of vertices with $d$ unmatched points and by
$$
pR=\sum_{i=1}^dR_i,
$$
the total number of unmatched points that are taken into account for the second selection.

In the same way as for Algorithm~\ref{alg:2}, we can derive expressions for the expected changes of the variables $r_i$.   For $0\le i \leq \lceil d/2 \rceil$, we have
\begin{align*}
\Delta (r_i)&=-2\frac{ir_i}{pL}\frac{ir_i}{pR}\\&-\frac{ir_i}{pL}\frac{pR-ir_i-(i+1)r_{i+1})}{pR}-\frac{pL-ir_i-(i+1)r_{i+1}}{pL}\frac{ir_i}{pR}\\
&+\frac{(i+1)r_{i+1}}{pL}\frac{nW-ir_i-(i+1)r_{i+1}}{pR}+\frac{pL-ir_i-(i+1)r_{i+1}}{pL}\frac{(i+1)r_{i+1}}{pR}\\&+2\frac{(i+1)r_{i+1}}{pL}\frac{(i+1)r_{i+1}}{pR},
\end{align*}
and for $\lceil d/2 \rceil < i \leq d$, we have
$$
\Delta (r_i)=-\frac{ir_i}{pR}+\frac{(i+1)r_{i+1}}{pR},
$$ 
and
$$
\Delta (r_d)=-\frac{dr_d}{pR}.
$$
By applying the differential equation method, these expected changes are translated to the  system of differential equations: 

\begin{align*}
    r'_i&=\frac{1}{pR\cdot pL}(2(ir_i)((i+1)r_{i+1})+(pR+pL)((i+1)r_{i+1}-ir_i)),&0\le i<\lceil d/2 \rceil \\
    r'_i&=\frac{1}{pR}((i+1)r_{i+1}-ir_i,&\lceil d/2 \rceil < i<d\\
    r'_d&=-\frac{1}{pR}dr_d. &
\end{align*}

The approximate solution to this system of differential equations at the stopping time  provides an asymptotically good approximation of the values of the variables on the output of Algorithm \ref{alg:3}. 
For the implementation, as before, we integrate the system of differential equations with the Runge-Kutta solver DOP853 in Python-3 with  $10^6$  steps. The system is initialized with the values obtained with the DEM for Algorithm~\ref{alg:2}.

\section{Results}\label{sec:results}

Table \ref{fig:results} summarizes the approximate upper and lower bounds for the vertex bisection width. In all of the cases, following  \cite{Kolesnik2014}, the displayed values give the proportion $\alpha$ of vertices in the red part which have neighbors in the other part, so that the value of the bisection is $\alpha (n/2)$.  The first column correspond to the DEM upper bound,  obtained by 
approximating the solution of the continuous differential equations generating by the changes of solution provided by the greedy algorithm~\ref{alg:2} (except for $d=3$). 
As mentioned before, it uses the Runge-Kuta solver DOP853  in Python, with a number $10^7$ of steps initialized with  $\epsilon=10^{-5}$, except for $d=9$ in which to avoid numerical errors we set $\epsilon=10^{-4}$.   
In Appendix~\ref{ape:B}, we provide the empirical results of the execution of Algorithm~\ref{alg:1}. From these results, in the second column, we present the \emph{experimental} upper bound for the vertex bisection width on random $d$-regular graphs, the highest minmum value attained in the executions of Algorithm~\ref{alg:1} on such graphs.
The third column  contains the lower bounds values obtained by Kolesnik and Wormald .

\begin{figure}[t]
\begin{center}
\begin{tabular}{|c|ccc|}
\hline
Degree $d$ & DEM  & Exper & LB \\
\hline
3  & $0.27964$ \cite{Lichev2020} & $0.30924$ & $0.14420$ \cite{Kolesnik2014}  \\
4  & $0.58103$ & $0.46552$ & $0.28966$ \cite{Kolesnik2014} \\
5  & $0.61018$ & $0.55903$ & $0.40859$ \cite{Kolesnik2014} \\
6  & $0.65693$ & $0.62588$ & $0.50190$ \cite{Kolesnik2014} \\
7  & $0.65640$ & $0.67865$ & $0.57466$ \cite{Kolesnik2014} \\
8  & $0.72031$ & $0.72051$ & $0.63178$ \cite{Kolesnik2014}\\
9  & $0.88097$ & $0.75354$ & $0.67716$ \cite{Kolesnik2014}\\
10 & $0.83769$ & $0.77800$ & $0.71371$ \cite{Kolesnik2014} \\
\hline
\end{tabular}
\end{center}
\caption{Values of $\alpha$ as upper bound via the DEM (except for $d=3$), as upper bound via empirical calculation and as lower bound, 
with degree $3\le d \le 10$.}\label{fig:results}
\end{figure}

Some comments are in order for an interpretation of the results. For $d=3$, the best upper and lower bounds obtained by Lichev and Mitsche \cite{Lichev2020} for the edge bisection width of random cubic graphs are 
$$
0.103295n\le \bw(G_{n,3})\le 0.139822n.
$$
The upper bound for the edge bisection width is also upper bound for vertex bisection width. Algorithm \ref{alg:1} on random cubic graphs gives an upper bound of $0.15782n$, while Algorithm \ref{alg:2} with DEM gives $0.24093n$, none of which beat the upper bound that can be obtained from the edge bisection width. This may be an indication that for random cubic graphs the structure of the bisection is closer to a matching than to the boundary of a ball. 

For $d>4$, the upper bounds on edge bisection for random $d$-regular graphs reported in \cite{Diaz2007}, do not provide too much information, as all of them are bigger than $0.5n$. For $d=4$, an upper bound for vertex bisection of $0.333 n$ is given in \cite{Diaz2003}. This gives an $\alpha$ value of 0.6666 which is improved by our DEM upper bound.

A second remark concerns the choice of $n$ in the range $10^5$ to $6\cdot 10^5$ for the size of graphs in our experiments. 
As mentioned before, the bisection width computed by Algorithm~\ref{alg:1} is not monotone in $n$. For instance, for $d=4$, the results of 
Algorithm~\ref{alg:1} are displayed in Table~\ref{tab:d4}. The selected range covers graph sizes that are large enough to be significant and that allow to carry out the computation. Besides, the range expands over sufficiently  many values of $n$ to capture the effects of the closeness of the size of the initial ball to $n/2$ for most of the considered values of $d$. Additional results for smaller values of $n$ are summarized in  Appendix~\ref{ape:B}.

The third remark concerns the values of the degrees up to $d=10$. The trivial upper bound for the minimum vertex bisection with is $n/2$, meaning that all vertices in one side of the bisection have a neighbour in the other side.  Indeed, for larger values of $d$ the lower bounds in \cite{Kolesnik2014} come close to $n/2$. For example, for $d=100$ the lower bound in \cite{Kolesnik2014} is $0.9785(n/2)$. This is also close to the expected value of a random bisection, showing very small variability among distinct bisections. We note that the expected vertex bisection of a random partition into two equal parts is $n/2(1-1/2^d)$ which tends to $n/2$ with $d\to \infty$. Therefore, the problem is meaningful only for a range of small values of the degree. Both algorithms presented here can be run for larger values of $d$. We have chosen to report only the range of degrees displayed in 
Table~\ref{fig:results}, which seem to us  the relevant ones, for which the upper bound obtained from the edge bisection width is meaningless and the trivial upper bound of $n/2$ can be significantly improved. 

A fourth remark  obviously concerns the gap between the lower and upper bounds for the vertex bisection width. The lower bound is obtained by a first moment method. However, the analysis is considerably more involved than the analogous application of the first moment method for the lower bound of the edge bisection width used by Bollob\'as \cite{Bollobas1988}. The union bound involved in the computation is likely to weaken the strength of the lower bounds, which we believe have room for improvement. On the other hand, the choice of a ball as a candidate for minimizing the vertex bisection width looks natural and we suspect that it may be closer to the optimal value.

As a final remark, we have concentrated on the vertex bisection problem instead of the general vertex isoperimetric problem. The latter asks for the minimum boundary of sets with given cardinality, other than $n/2$.  Both algorithms presented here are suited to analyze this more general problem, by setting the stopping time at $cn$ for different values of $c$ other than $1/2$. A Python implementation of the algorithms is available on demand.

\section{Conclusions}\label{sec:conclusions}

The edge bisection problem, or more general the edge isoperimetric problem, has been intensively studied in the literature and has a wide range of applications. 
One of the reasons is that there are available tools, from randomized algorithms to spectral techniques, which make the problem more  approachable than its related 
vertex bisection problem, 
or the more general vertex isoperimetric problem. For constant degree, most of the applications to computer science and other areas, bounds on edge isoperimetry are usually enough even if vertex isoperimetry is concerned. However, from the theoretical point of view, it is  a natural and important question to understand the behaviour of the two parameters.

The vertex isoperimetric problem is solved for a few classes of graphs. A particularly  significant one is the solutions by Harper~\cite{Harper1964,Harper1966}  for the vertex and edge isoperimetric problems on the hypercube. Interestingly, the minimizers for both problems in the hypercube are structurally very different. This prompted us to investigate the vertex bisection  problem for random regular graphs.

For fixed small degree, the best bounds for the edge vertex bisection of random regular graphs have been obtained by Diaz, Serna and Wormald \cite{Diaz2007} by analyzing a greedy algorithm with the differential equations method. The greedy algorithm used in \cite{Diaz2007} does not provide good vertex bisectors, indicating that the minimizing bisectors for the case of edge cuts and vertex cuts have distinct structural properties. In this paper we consider a different randomized greedy algorithm to obtain upper bounds on the vertex bisection width of random regular graphs, which is also analyzed with the differential equation method. The results are compared with the lower bounds for this parameter obtained by Kolesnik and Wormald \cite{Kolesnik2014} and also with experimental results with a simplified version of the randomized greedy algorithm. To the best of our knowledge our algorithm provides the first upper bounds for the vertex bisection problem of random regular graphs. The gaps between lower and upper bounds certainly leave room for improvement, which is the more relevant natural question left open.




\appendix
\newpage
\section{The evolution of the Balls\label{ape:A}}
The following plots depict the sizes of the balls ($y$-coordinate) as function of the number of nodes ($x$-coordinate),  for the different values of $d$. In the plots, B0, corresponds to the size of the ball $B(x_0,r_0)$. Recall that $r_0$ is selected in such a way that $|B(x_0,r_0+2)|\geq n$. B1 is the size of $B(x_0,r_0+1)$ and  B2 that of $B(x_0,r_0+2)$.

\begin{center}
\includegraphics[width=6cm]{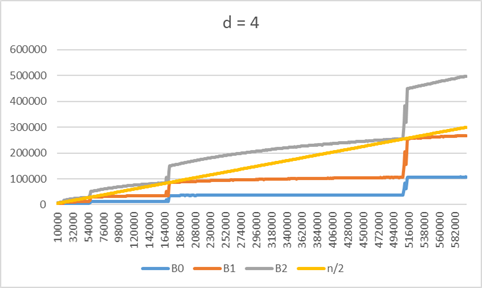}\hspace{5mm}\includegraphics[width=6cm]{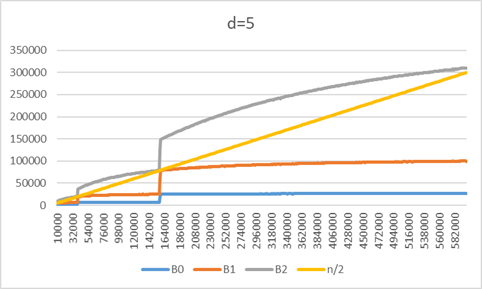}

\includegraphics[width=6cm]{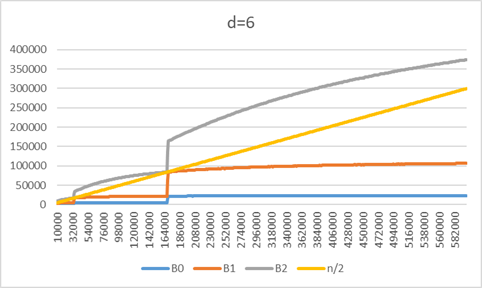}\hspace{5mm}\includegraphics[width=6cm]{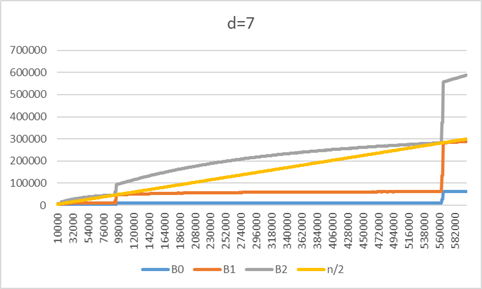}

\includegraphics[width=6cm]{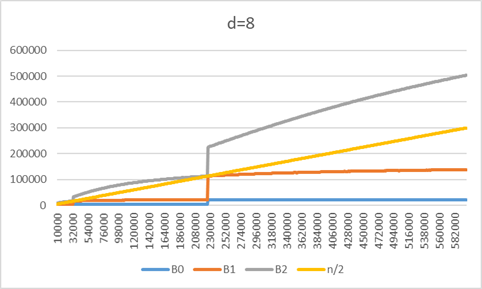}\hspace{5mm}\includegraphics[width=6cm]{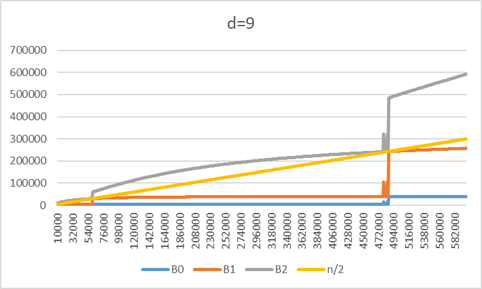}

\includegraphics[width=6cm]{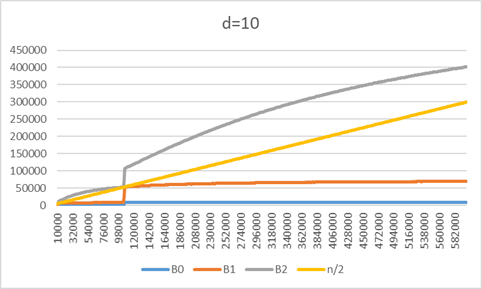}
\end{center}
\newpage
\section{Experimental results\label{ape:B}}

In the following tables, we present the computed values of the vertex bisections obtained after running  Algorithm~\ref{alg:1} rounded to the fifth digit.  

For each value of $d$ in the interval $[4,10]$, we generated 5 random graphs of each size and executed the algorithm  5 times on each graph.  In the tables, the runs are identified as $r_0$ to $r_5$. We also show the average, maximum and minimum vertex bisection of each graph. We highlight in red the highest minimum vertex bisection and in blue, the same value for each graph size. 
For small values of $n$, we omit the results of  each run and provide the average, max and min values for the graph having the worst seen vertex bisection.

\begin{table}[h]
\small
\centering
\begin{tabular}{|l|l|l|l|l||l|l|l|l|}
\hline
$d$ & $n$ & Avg & Max & Min &  $n$ & Avg & Max & Min \\
\hline
4& 1000 & 0.47480 &	0.48800 & \tred{0.47000} & 4000 & 0.46150 & 0.46300 & 0.45950\\
4& 2000 & 0.46540 &	0.47700	& 0.45700 &  5000 & 0.46008 & 0.46360 & 0.45680\\
4& 3000 & 0.46360 &	0.46600 & 0.46133 &  6000 & 0.46080 & 0.46233 & 0,45933\\
\hline
5& 1000 & 0.56680 & 0.57000 & \tred{0.56200} & 4000 & 0.56170 & 0,56500 & 0.55900\\
5& 2000 & 0.55900 & 0.56300 & 0.55500 & 5000 & 0.55920 & 0.56440 & 0.55600\\
5& 3000 & 0.56453 & 0.56733 & 0.55933 & 6000 & 0.55833 & 0.56233 & 0.55500\\
\hline
6& 1000 & 0.63400 & 0.64000 & \tred{0.62800} & 4000 & 0.62600 & 0.63050 & 0.62150\\
6& 2000 & 0.62920 & 0.63200 & 0.62500 & 5000 & 0.62848 & 0.63120 & 0.62520\\
6& 3000 & 0.62787 & 0.63000	& 0.62600 & 6000 & 0.62433 & 0.62667 & 0.62300\\
\hline
7& 1000 & 0.68240 & 0.68600 & 0.68000 & 4000 & 0.67990 & 0.68400 & 0.67650\\
7& 2000 & 0.68300 & 0.68500 & \tred{0.68100} & 5000 & 0.67952 & 0.68080 & 0.67680\\
7& 3000 & 0.68173 & 0.68733 & 0.67733 & 6000 & 0.68093 & 0.68567 & 0.67867\\
\hline
8& 1000 & 0.72440 & 0.72800 & 0.72000 & 4000 & 0.72220 & 0.72650 & 0.71900\\
8& 2000 & 0.72020 & 0.72200 & 0.71800 & 5000 & 0.72440 & 0.72760 & 0.72120\\
8& 3000 & 0.72400 & 0.72800 & \tred{0.72133} & 6000 & 0.72153 & 0.72333 & 0.72033\\
\hline
9& 1000 & 0.75640 & 0.76600 & \tred{0.75400} & 4000 & 0.75490 & 0.75600 & 0.75300\\
9& 2000 & 0.75460 & 0.75700 & 0.75300 & 5000 & 0.75504 & 0.75640 & \tred{0.75400}\\
9& 3000 & 0.75520 & 0.75667 & \tred{0.75400} & 6000 & 0.75453 & 0.75733 & 0.75233\\
\hline
10& 1000& 0.78360 & 0.78600 & 0.77800 & 4000& 0.78000 & 0.78250 & 0.77750\\
10& 2000& 0.78100 & 0.78300 & \tred{0.78000} & 5000& 0.77944 & 0.78080 & 0,77760\\
10& 3000& 0.77960 & 0.78000 & 0.77867 & 6000& 0.77867 & 0.77967 & 0.77800\\
\hline
\end{tabular}
\caption{Experimental results of Algorithm~\ref{alg:1} for small values of $n$.}
\end{table}
\newpage
\begin{table}[h]
\small
\begin{tabular}{|l|l|l|l|l|l|l|l|l|l|l||l|l|l|}
\hline
$n$&$r_0$&$r_1$&$r_2$&$r_3$&$r_4$&Avg&Max &Min\\
\hline
100000&0.30882 &  0.31014 &  0.31050 &  0.30812 &  0.30810 & 0.30914 & 0.31050 & 0.30810\\
100000&0.30918 &  0.30934 &  0.30846 &  0.30946 &  0.30756 & 0.30880 & 0.30946 & 0.30756\\
100000&0.30902 &  0.30986 &  0.30974 &  0.31018 &  0.31048 & 0.30986 & 0.31048 & 0.30902\\
100000&0.31162 &  0.30828 &  0.31308 &  0.30914 &  0.31108 & 0.31064 & 0.31308 & 0.30828\\
100000&0.31060 &  0.30938 &  0.30924 &  0.31104 &  0.30978 & 0.31001 & 0.31104 & \tblue{0.30924}\\ \hline
200000&0.30825 &  0.3093  &  0.30919 &  0.30947 &  0.30911 & 0.30906 & 0.30947 & 0.30825\\
200000&0.30951 &  0.3086  &  0.31078 &  0.31013 &  0.30771 & 0.30935 & 0.31078 & 0.30771\\
200000&0.31004 &  0.30908 &  0.30829 &  0.31067 &  0.30794 & 0.30920 & 0.31067 & 0.30794\\
200000&0.30909 &  0.31003 &  0.31075 &  0.30900 &  0.31016 & 0.30981 & 0.31075 & \tblue{0.30900}\\
200000&0.30814 &  0.31100 &  0.30854 &  0.31038 &  0.31056 & 0.30972 & 0.31100 & 0.30814\\ \hline
300000&0.30865 &  0.30953 &  0.31051 &  0.30949 &  0.30899 & 0.30943 & 0.31051 & 0.30865\\
300000&0.30975 &  0.30932 &  0.30872 &  0.30931 &  0.30888 & 0.30920 & 0.30975 & \tblue{0.30872}\\
300000&0.30907 &  0.30932 &  0.30925 &  0.30779 &  0.30991 & 0.30907 & 0.30991 & 0.30779\\
300000&0.30985 &  0.30893 &  0.31051 &  0.30953 &  0.30823 & 0.30941 & 0.31051 & 0.30823\\
300000&0.30947 &  0.31076 &  0.30833 &  0.30904 &  0.30926 & 0.30937 & 0.31076 & 0.30833\\ \hline
400000&0.30944 &  0.30931 &  0.30937 &  0.31012 &  0.30909 & 0.30946 & 0.31012 & 0.30909\\ 
400000&0.30838 &  0.30861 &  0.30983 &  0.31019 &  0.30889 & 0.30918 & 0.31019 & 0.30838\\
400000&0.30931 &  0.30926 &  0.30922 &  0.30967 &  0.30962 & 0.30941 & 0.30967 & \tblue{0.30922}\\
400000&0.30903 &  0.30900 &  0.31074 &  0.30958 &  0.31002 & 0.30967 & 0.31074 & 0.30900\\
400000&0.30950 &  0.30974 &  0.30830 &  0.31060 &  0.30947 & 0.30952 & 0.31060 & 0.30830\\ \hline
500000&0.30965 &  0.30828 &  0.30990 &  0.30961 &  0.30902 & 0.30929 & 0.30990 & 0.30828\\
500000&0.31000 &  0.30988 &  0.30994 &  0.30904 &  0.30838 & 0.30945 & 0.31000 & 0.30838\\
500000&0.30942 &  0.30939 &  0.31010 &  0.30940 &  0.31021 & 0.30970 & 0.31021 & \tblue{0.30939}\\
500000&0.30941 &  0.31036 &  0.30874 &  0.30934 &  0.30958 & 0.30949 & 0.31036 & 0.30874\\
500000&0.30805 &  0.30856 &  0.30897 &  0.30964 &  0.30970 & 0.30898 & 0.30970 & 0.30805\\\hline
600000&0.30902 &  0.30922 &  0.30989 &  0.30973 &  0.30990 & 0.30955 & 0.30990 & 0.30902\\
600000&0.31028 &  0.30990 &  0.30902 &  0.30947 &  0.31004 & 0.30974 & 0.31028 & 0.30902\\
600000&0.30928 &  0.31038 &  0.30864 &  0.30962 &  0.30937 & 0.30946 & 0.31038 & 0.30864\\
600000&0.30958 &  0.30990 &  0.31075 &  0.30976 &  0.30967 & 0.30993 & 0.31075 & \tred{0.30958}\\
600000&0.30945 &  0.30881 &  0.30863 &  0.30983 &  0.30898 & 0.30914 & 0.30983 & 0.30863\\
\hline
\end{tabular}
\caption{Experimental results of Algorithm~\ref{alg:1}, for $d=3$.}
\end{table}

\begin{table}[h]
\small
\begin{tabular}{|l|l|l|l|l|l|l|l|l|l|l||l|l|l|}
\hline
$n$&$r_0$&$r_1$&$r_2$&$r_3$&$r_4$&Avg&Max &Min\\
\hline
100000&0.46270&0.46206&0.46048&0.46104&0.45950&0.46116&0.46270&0.45950\\
100000&0.46190&0.46108&0.46246&0.45972&0.46142&0.46132&0.46246&0.45972\\
100000&0.46204&0.46198&0.46154&0.46106&0.46072&0.46147&0.46204&\tblue{0.46072}\\
100000&0.45968&0.46082&0.46166&0.45920&0.46096&0.46046&0.46166&0.45920\\
100000&0.46270&0.46426&0.46112&0.46164&0.45928&{0.46180}&0.46426&0.45928\\ \hline
200000&0.46535&0.46573&0.46584&0.46698&0.46416&0.46561&0.46698&0.46416\\
200000&0.46647&0.46629&0.46552&0.46702&0.46620&{0.46630}&0.46702&\tred{0.46552}\\
200000&0.46454&0.46727&0.46649&0.46642&0.46588&0.46612&0.46727&0.46454\\
200000&0.46687&0.46550&0.46653&0.46419&0.46446&0.46551&0.46687&0.46419\\
200000&0.46527&0.46521&0.46631&0.46367&0.46708&0.46551&0.46708&0.46367\\ \hline
300000&0.46076&0.46080&0.46293&0.46062&0.46041&0.46111&0.46293&0.46041\\
300000&0.46143&0.45971&0.46169&0.46154&0.46090&0.46105&0.46169&0.45971\\
300000&0.46085&0.46035&0.46136&0.46140&0.46135&0.46106&0.46140&0.46035\\
300000&0.46111&0.46061&0.46082&0.46127&0.46112&0.46099&0.46127&0.46061\\
300000&0.46149&0.46145&0.46191&0.46126&0.46065&{0.46135}&0.46191&\tblue{0.46065}\\ \hline
400000&0.45826&0.45841&0.45873&0.45890&0.45849&0.45856&0.45890&0.45826\\
400000&0.45788&0.45902&0.45823&0.45821&0.45946&0.45856&0.45946&0.45788\\
400000&0.45814&0.45855&0.45783&0.45900&0.45881&0.45846&0.45900&0.45783\\
400000&0.45884&0.45929&0.45929&0.45901&0.45906&{0.45910}&0.45929&\tblue{0.45884}\\
400000&0.45861&0.45916&0.45956&0.45888&0.45858&0.45896&0.45956&0.45858\\ \hline
500000&0.45882&0.45930&0.45794&0.45813&0.45880&0.45860&0.45930&0.45794\\
500000&0.45952&0.45899&0.45800&0.45856&0.45821&0.45866&0.45952&0.45800\\
500000&0.45833&0.45983&0.45933&0.45906&0.45869&{0.45905}&0.45983&0.45833\\
500000&0.45890&0.45908&0.45883&0.45836&0.45914&0.45886&0.45914&0.45836\\
500000&0.45928&0.45867&0.45911&0.45841&0.45850&0.45879&0.45928&\tblue{0.45841}\\ \hline
600000&0.46539&0.46567&0.46519&0.46570&0.46540&0.46547&0.46570&0.46519\\
600000&0.46555&0.46517&0.46556&0.46573&0.46590&0.46558&0.46590&0.46517\\
600000&0.46611&0.46600&0.46532&0.46589&0.46624&0.46591&0.46624&0.46532\\
600000&0.46540&0.46590&0.46540&0.46635&0.46605&0.46582&0.46635&\tblue{0.46540}\\
600000&0.46523&0.46638&0.46563&0.46632&0.46614&{0.46594}&0.46638&0.46523\\
\hline
\end{tabular}
\caption{Experimental results of Algorithm~\ref{alg:1}, for $d=4$.}
\end{table}

\begin{table}[h]
\small
\begin{tabular}{|l|l|l|l|l|l|l|l|l|l|l||l|l|l|}
\hline
$n$&$r_0$&$r_1$&$r_2$&$r_3$&$r_4$&Avg&Max &Min\\
\hline
100000 & 0.55856 & 0.55482 & 0.55674 & 0.55654 & 0.55606 & 0.55654 & 0.55856 & 0.55482\\
100000  & 0.55614 & 0.55488 & 0.55604 & 0.55676 & 0.55652 & 0.55607 & 0.55676 & 0.55488\\
100000 & 0.55642 & 0.55522 & 0.55622 & 0.55866 & 0.55708 & {0.55672} & 0.55866 & 0.55522\\
100000 & 0.55600 & 0.55546 & 0.55802 & 0.55606 & 0.55710 & 0.55653 & 0.55802 & 0.55546\\
100000 & 0.55648 & 0.55626 & 0.55692 & 0.55616 & 0.55744 & 0.55665 & 0.55744 & \tblue{0.55616}\\ \hline
200000 & 0.56019 & 0.55838 & 0.55864 & 0.55914 & 0.55929 & 0.55913 & 0.56019 & 0.55838\\
200000 & 0.55961 & 0.55907 & 0.55936 & 0.55903 & 0.55998 & {0.55941} & 0.55998 & \tred{0.55903}\\
200000 & 0.55915 & 0.55881 & 0.55874 & 0.55920 & 0.55941 & 0.55906 & 0.55941 & 0.55874\\
200000 & 0.55897 & 0.55857 & 0.55939 & 0.55874 & 0.56079 & 0.55929 & 0.56079 & 0.55857\\
200000 & 0.55941 & 0.56002 & 0.55881 & 0.55868 & 0.55951 & 0.55929 & 0.56002 & 0.55868\\ \hline
300000 & 0.55817 & 0.55749 & 0.55803 & 0.55745 & 0.55693 & 0.55761 & 0.55817 & 0.55693\\
300000 & 0.55700 & 0.55735 & 0.55733 & 0.55743 & 0.55757 & 0.55734 & 0.55757 & 0.55700\\
300000 & 0.55740 & 0.55729 & 0.55787 & 0.55799 & 0.55736 & 0.55758 & 0.55799 & \tblue{0.55729}\\
300000 & 0.55704 & 0.55790 & 0.55741 & 0.55703 & 0.55767 & 0.55741 & 0.55790 & 0.55703\\
300000 & 0.55697 & 0.55777 & 0.55724 & 0.55789 & 0.55832 & {0.55764} & 0.55832 & 0.55697\\ \hline
400000 & 0.55574 & 0.55636 & 0.55708 & 0.55633 & 0.55663 & 0.55643 & 0.55708 & 0.55574\\
400000 & 0.55735 & 0.55679 & 0.55613 & 0.55627 & 0.55625 & 0.55656 & 0.55735 & \tblue{0.55613}\\
400000 & 0.55716 & 0.55660 & 0.55716 & 0.55709 & 0.55600 & {0.55680} & 0.55716 & 0.55600\\
400000 & 0.55644 & 0.55556 & 0.55676 & 0.55668 & 0.55669 & 0.55643 & 0.55676 & 0.55556\\
400000 & 0.55717 & 0.55619 & 0.55646 & 0.55686 & 0.55598 & 0.55653 & 0.55717 & 0.55598\\ \hline
500000 & 0.55618 & 0.55634 & 0.55599 & 0.55627 & 0.55637 & {0.55623} & 0.55637 & \tblue{0.55599}\\
500000 & 0.55560 & 0.55633 & 0.55659 & 0.55615 & 0.55633 & 0.55620 & 0.55659 & 0.55560\\
500000 & 0.55667 & 0.55641 & 0.55539 & 0.55555 & 0.55594 & 0.55599 & 0.55667 & 0.55539\\
500000 & 0.55608 & 0.55592 & 0.55536 & 0.55627 & 0.55604 & 0.55593 & 0.55627 & 0.55536\\
500000 & 0.55622 & 0.55589 & 0.55563 & 0.55658 & 0.55549 & 0.55596 & 0.55658 & 0.55549\\ \hline
600000 & 0.55573 & 0.55561 & 0.55525 & 0.55591 & 0.55518 & 0.55553 & 0.55591 & 0.55518\\
600000 & 0.55586 & 0.55524 & 0.55593 & 0.55525 & 0.55573 & 0.55560 & 0.55593 & 0.55524\\
600000 & 0.55520 & 0.55529 & 0.55530 & 0.55589 & 0.55591 & 0.55552 & 0.55591 & 0.55520\\
600000 & 0.55632 & 0.55534 & 0.55597 & 0.55538 & 0.55564 & {0.55573} & 0.55632 & \tblue{0.55534}\\
600000 & 0.55510 & 0.55536 & 0.55612 & 0.55547 & 0.55561 & 0.55553 & 0.55612 & 0.55510\\
\hline
\end{tabular}
\caption{Experimental results of Algorithm~\ref{alg:1}, for $d=5$.}
\end{table}

\begin{table}
\small
\begin{tabular}{|l|l|l|l|l|l|l|l|l|l|l||l|l|l|}
\hline
$n$&$r_0$&$r_1$&$r_2$&$r_3$&$r_4$&Avg&Max &Min\\
\hline
100000 & 0.62408 & 0.62328 & 0.62372 & 0.62370 & 0.62486 & 0.62393 & 0.62486 & 0.62328\\
100000 & 0.62522 & 0.62402 & 0.62564 & 0.62434 & 0.62452 & 0.62475 & 0.62564 & \tblue{0.62402}\\
100000 & 0.62384 & 0.62464 & 0.62448 & 0.62522 & 0.62492 & 0.62462 & 0.62522 & 0.62384\\
100000 & 0.62356 & 0.62564 & 0.62552 & 0.62524 & 0.62436 & {0.62486} & 0.62564 & 0.62356\\
100000 & 0.62436 & 0.62432 & 0.62442 & 0.62282 & 0.62416 & 0.62402 & 0.62442 & 0.62282\\ \hline
200000 & 0.62647 & 0.62691 & 0.62664 & 0.62602 & 0.62527 & 0.62626 & 0.62691 & 0.62527\\
200000 & 0.62605 & 0.62558 & 0.62717 & 0.62621 & 0.62633 & 0.62627 & 0.62717 & 0.62558\\
200000 & 0.62581 & 0.62690 & 0.62710 & 0.62649 & 0.62646 & {0.62655} & 0.62710 & 0.62581\\
200000 & 0.62587 & 0.62624 & 0.62648 & 0.62647 & 0.62603 & 0.62622 & 0.62648 & 0.62587\\
200000 & 0.62631 & 0.62688 & 0.62628 & 0.62658 & 0.62588 & 0.62639 & 0.62688 & \tred{0.62588}\\ \hline
300000 & 0.62447 & 0.62466 & 0.62501 & 0.62507 & 0.62541 & 0.62493 & 0.62541 & 0.62447\\
300000 & 0.62491 & 0.62491 & 0.62537 & 0.62523 & 0.62509 & 0.62510 & 0.62537 & 0.62491\\
300000 & 0.62483 & 0.62497 & 0.62463 & 0.62440 & 0.62489 & 0.62475 & 0.62497 & 0.62440\\
300000 & 0.62559 & 0.62497 & 0.62514 & 0.62519 & 0.62595 & 0.62537 & 0.62595 & \tblue{0.62497}\\ 
300000 & 0.62487 & 0.62537 & 0.62523 & 0.62539 & 0.62610 & {0.62539} & 0.62610 & 0.62487\\ \hline
400000 & 0.62483 & 0.62454 & 0.62389 & 0.62435 & 0.62475 & 0.62447 & 0.62483 & 0.62389\\
400000 & 0.62469 & 0.62457 & 0.62441 & 0.62443 & 0.62481 & 0.62458 & 0.62481 & 0.62441\\
400000 & 0.62384 & 0.62503 & 0.62509 & 0.62498 & 0.62415 & 0.62462 & 0.62509 & 0.62384\\
400000 & 0.62443 & 0.62392 & 0.62475 & 0.62435 & 0.62478 & 0.62445 & 0.62478 & 0.62392\\
400000 & 0.62471 & 0.62467 & 0.62488 & 0.62508 & 0.62485 & {0.62484} & 0.62508 & 0.62467\\ \hline
500000 & 0.62425 & 0.62447 & 0.62390 & 0.62460 & 0.62428 & 0.62430 & 0.62460 & 0.62390\\ 
500000 & 0.62448 & 0.62438 & 0.62461 & 0.62500 & 0.62448 & {0.62459} & 0.62500 & \tblue{0.62438}\\
500000 & 0.62442 & 0.62438 & 0.62473 & 0.62432 & 0.62404 & 0.62438 & 0.62473 & 0.62404\\
500000 & 0.62472 & 0.62447 & 0.62448 & 0.62433 & 0.62454 & 0.62451 & 0.62472 & 0.62433\\
500000 & 0.62417 & 0.62432 & 0.62418 & 0.62441 & 0.62491 & 0.62440 & 0.62491 & 0.62417\\ \hline
600000 & 0.62422 & 0.62483 & 0.62406 & 0.62411 & 0.62407 & {0.62426} & 0.62483 & \tblue{0.62406}\\
600000 & 0.62380 & 0.62376 & 0.62404 & 0.62353 & 0.62422 & 0.62387 & 0.62422 & 0.62353\\
600000 & 0.62449 & 0.62469 & 0.62438 & 0.62376 & 0.62367 & 0.62420 & 0.62469 & 0.62367\\
600000 & 0.62402 & 0.62424 & 0.62404 & 0.62402 & 0.62483 & 0.62423 & 0.62483 & 0.62402\\
600000 & 0.62432 & 0.62416 & 0.62416 & 0.62379 & 0.62414 & 0.62412 & 0.62432 & 0.62379\\

\hline
\end{tabular}
\caption{Experimental results of Algorithm~\ref{alg:1} for $d=6$.}
\end{table}

\begin{table}
\small
\begin{tabular}{|l|l|l|l|l|l|l|l|l|l|l||l|l|l|}
\hline
$n$&$r_0$&$r_1$&$r_2$&$r_3$&$r_4$&Avg&Max &Min\\
\hline
100000 & 0.67908 & 0.67916 & 0.67784 & 0.67860 & 0.67780 & 0.67850 & 0.67916 & 0.67780\\
100000 & 0.67816 & 0.67956 & 0.67910 & 0.67936 & 0.67942 & 0.67912 & 0.67956 & 0.67816\\
100000 & 0.67888 & 0.68000 & 0.67962 & 0.67904 & 0.67854 & {0.67922} & 0.68000 & \tred{0.67854}\\
100000 & 0.67908 & 0.67814 & 0.67852 & 0.67882 & 0.67820 & 0.67855 & 0.67908 & 0.67814\\
100000 & 0.67878 & 0.67988 & 0.67808 & 0.67976 & 0.67900 & 0.67910 & 0.67988 & 0.67808\\ \hline
200000 & 0.67725 & 0.67672 & 0.67692 & 0.67663 & 0.67638 & 0.67678 & 0.67725 & 0.67638\\
200000 & 0.67649 & 0.67728 & 0.67660 & 0.67693 & 0.67644 & 0.67675 & 0.67728 & 0.67644\\
200000 & 0.67775 & 0.67729 & 0.67656 & 0.67672 & 0.67753 & {0.67717} & 0.67775 & \tblue{0.67656}\\
200000 & 0.67695 & 0.67856 & 0.67659 & 0.67672 & 0.67656 & 0.67708 & 0.67856 & 0.67656\\
200000 & 0.67714 & 0.67668 & 0.67753 & 0.67621 & 0.67690 & 0.67689 & 0.67753 & 0.67621\\ \hline
300000 & 0.67643 & 0.67675 & 0.67735 & 0.67677 & 0.67711 & {0.67688} & 0.67735 & 0.67643\\
300000 & 0.67645 & 0.67728 & 0.67699 & 0.67682 & 0.67635 & 0.67678 & 0.67728 & 0.67635\\
300000 & 0.67703 & 0.67638 & 0.67665 & 0.67657 & 0.67693 & 0.67671 & 0.67703 & 0.67638\\
300000 & 0.67693 & 0.67645 & 0.67693 & 0.67633 & 0.67763 & 0.67685 & 0.67763 & 0.67633\\
300000 & 0.67631 & 0.67619 & 0.67709 & 0.67662 & 0.67637 & 0.67652 & 0.67709 & 0.67619\\ \hline
400000 & 0.67627 & 0.67654 & 0.67678 & 0.67721 & 0.67677 & {0.67671} & 0.67721 & \tblue{0.67627}\\
400000 & 0.67662 & 0.67592 & 0.67663 & 0.67671 & 0.67638 & 0.67645 & 0.67671 & 0.67592\\
400000 & 0.67591 & 0.67617 & 0.67690 & 0.67637 & 0.67645 & 0.67636 & 0.67690 & 0.67591\\
400000 & 0.67681 & 0.67670 & 0.67672 & 0.67633 & 0.67589 & 0.67649 & 0.67681 & 0.67589\\
400000 & 0.67661 & 0.67682 & 0.67642 & 0.67609 & 0.67650 & 0.67649 & 0.67682 & 0.67609\\ \hline
500000 & 0.67662 & 0.67670 & 0.67625 & 0.67636 & 0.67673 & {0.67653} & 0.67673 & \tblue{0.67625}\\
500000 & 0.67666 & 0.67656 & 0.67631 & 0.67654 & 0.67588 & 0.67639 & 0.67666 & 0.67588\\
500000 & 0.67640 & 0.67648 & 0.67595 & 0.67652 & 0.67625 & 0.67632 & 0.67652 & 0.67595\\
500000 & 0.67679 & 0.67641 & 0.67624 & 0.67624 & 0.67652 & 0.67644 & 0.67679 & 0.67624\\
500000 & 0.67640 & 0.67658 & 0.67623 & 0.67674 & 0.67614 & 0.67642 & 0.67674 & 0.67614\\ \hline
600000 & 0.67840 & 0.67863 & 0.67865 & 0.67867 & 0.67873 & 0.67861 & 0.67873 & 0.67840\\
600000 & 0.67807 & 0.67841 & 0.67884 & 0.67852 & 0.67859 & 0.67849 & 0.67884 & 0.67807\\
600000 & 0.67826 & 0.67880 & 0.67865 & 0.67892 & 0.67896 & {0.67872} & 0.67896 & 0.67826  \\
600000 & 0.67855 & 0.67890 & 0.67842 & 0.67821 & 0.67879 & 0.67857 & 0.67890 & 0.67821 \\
600000 & 0.67869 & 0.67872 & 0.67872 & 0.67867 & 0.67865 & 0.67869 & 0.67872 & \tblue{0.67865} \\
\hline
\end{tabular}
\caption{Experimental results of Algorithm~\ref{alg:1} for $d=7$.}
\end{table}

\begin{table}
\small
\begin{tabular}{|l|l|l|l|l|l|l|l|l|l|l||l|l|l|}
\hline
$n$&$r_0$&$r_1$&$r_2$&$r_3$&$r_4$&Avg&Max &Min\\
\hline
100000 & 0.72096 & 0.71836 & 0.72024 & 0.71960 & 0.71926 & 0.71968 & 0.72096 & 0.71836\\
100000 & 0.72028 & 0.72002 & 0.71910 & 0.71934 & 0.71950 & 0.71965 & 0.72028 & 0.71910\\
100000 & 0.71956 & 0.71920 & 0.72024 & 0.71908 & 0.72098 & {0.71981} & 0.72098 & 0.71908\\
100000 & 0.71972 & 0.71926 & 0.71938 & 0.71966 & 0.71954 & 0.71951 & 0.71972 & 0.71926\\
100000 & 0.71958 & 0.71940 & 0.71932 & 0.71956 & 0.72048 & 0.71967 & 0.72048 & \tblue{0.71932}\\ \hline
200000 & 0.71921 & 0.71836 & 0.71965 & 0.71978 & 0.71920 & 0.71924 & 0.71978 & 0.71836\\
200000 & 0.71943 & 0.71886 & 0.71948 & 0.71907 & 0.71937 & 0.71924 & 0.71948 & 0.71886\\
200000 & 0.71921 & 0.71965 & 0.71992 & 0.71954 & 0.71969 & {0.71960} & 0.71992 & \tblue{0.71921}\\
200000 & 0.71894 & 0.71955 & 0.71973 & 0.71952 & 0.71918 & 0.71938 & 0.71973 & 0.71894\\
200000 & 0.71939 & 0.71848 & 0.71942 & 0.71976 & 0.71960 & 0.71933 & 0.71976 & 0.71848\\ \hline
300000 & 0.72090 & 0.72050 & 0.72036 & 0.72111 & 0.72041 & 0.72066 & 0.72111 & 0.72036\\
300000 & 0.72079 & 0.72099 & 0.72051 & 0.72079 & 0.72077 & {0.72077} & 0.72099 & \tred{0.72051}\\
300000 & 0.72051 & 0.72076 & 0.72032 & 0.72076 & 0.72000 & 0.72047 & 0.72076 & 0.72000\\
300000 & 0.72043 & 0.72038 & 0.72052 & 0.72023 & 0.72013 & 0.72034 & 0.72052 & 0.72013\\
300000 & 0.72028 & 0.72043 & 0.72079 & 0.72067 & 0.72065 & 0.72057 & 0.72079 & 0.72028\\ \hline
400000 & 0.71964 & 0.71969 & 0.72048 & 0.71968 & 0.72017 & 0.71993 & 0.72048 & 0.71964\\
400000 & 0.71974 & 0.71987 & 0.72020 & 0.72062 & 0.72002 & {0.72009} & 0.72062 & 0.71974\\
400000 & 0.72015 & 0.71991 & 0.72001 & 0.72059 & 0.71945 & 0.72002 & 0.72059 & 0.71945\\
400000 & 0.72006 & 0.72022 & 0.72024 & 0.71998 & 0.71964 & 0.72003 & 0.72024 & 0.71964\\
400000 & 0.72047 & 0.72002 & 0.71995 & 0.72017 & 0.71987 & 0.72009 & 0.72047 & \tblue{0.71987}\\ \hline
500000 & 0.71994 & 0.71979 & 0.71970 & 0.71983 & 0.71984 & 0.71982 & 0.71994 & \tblue{0.71970}\\
500000 & 0.72009 & 0.72026 & 0.71978 & 0.71965 & 0.71974 & {0.71990} & 0.72026 & 0.71965\\
500000 & 0.71966 & 0.72000 & 0.71991 & 0.71982 & 0.72008 & 0.71990 & 0.72008 & 0.71966\\
500000 & 0.71966 & 0.71950 & 0.71956 & 0.71953 & 0.71982 & 0.71961 & 0.71982 & 0.71950\\
500000 & 0.71953 & 0.71945 & 0.71990 & 0.72046 & 0.71972 & 0.71981 & 0.72046 & 0.71945\\ \hline
600000 & 0.71971 & 0.72008 & 0.72004 & 0.71969 & 0.71968 & {0.71984} & 0.72008 & \tblue{0.71968}\\
600000 & 0.71965 & 0.71996 & 0.71950 & 0.71966 & 0.71967 & 0.71969 & 0.71996 & 0.71950\\
600000 & 0.71943 & 0.71968 & 0.71952 & 0.71987 & 0.71950 & 0.71960 & 0.71987 & 0.71943\\
600000 & 0.72001 & 0.71962 & 0.71965 & 0.71960 & 0.71962 & 0.71970 & 0.72001 & 0.71960\\
600000 & 0.71951 & 0.71962 & 0.71977 & 0.71972 & 0.71998 & 0.71972 & 0.71998 & 0.71951\\
\hline
\end{tabular}
\caption{Experimental results of Algorithm~\ref{alg:1} for $d=8$.}
\end{table}

\begin{table}
\small
\begin{tabular}{|l|l|l|l|l|l|l|l|l|l|l||l|l|l|}
\hline
$n$&$r_0$&$r_1$&$r_2$&$r_3$&$r_4$&Avg&Max &Min\\
\hline
100000 & 0.75212 & 0.75228 & 0.75282 & 0.75272 & 0.75264 & 0.75252 & 0.75282 & \tblue{0.75212}\\
100000 & 0.75272 & 0.75186 & 0.75222 & 0.75142 & 0.75350 & 0.75234 & 0.75350 & 0.75142\\
100000 & 0.75190 & 0.75218 & 0.75322 & 0.75270 & 0.75236 & 0.75247 & 0.75322 & 0.75190\\
100000 & 0.75204 & 0.75320 & 0.75336 & 0.75290 & 0.75324 & {0.75295} & 0.75336 & 0.75204\\
100000 & 0.75336 & 0.75342 & 0.75308 & 0.75194 & 0.75286 & 0.75293 & 0.75342 & 0.75194\\ \hline
200000 & 0.75242 & 0.75252 & 0.75261 & 0.75196 & 0.75213 & 0.75233 & 0.75261 & 0.75196\\
200000 & 0.75257 & 0.75247 & 0.75256 & 0.75259 & 0.75250 & {0.75254} & 0.75259 & \tblue{0.75247}\\
200000 & 0.75213 & 0.75250 & 0.75211 & 0.75206 & 0.75287 & 0.75233 & 0.75287 & 0.75206\\
200000 & 0.75270 & 0.75227 & 0.75240 & 0.75274 & 0.75215 & 0.75245 & 0.75274 & 0.75215\\
200000 & 0.75260 & 0.75189 & 0.75213 & 0.75220 & 0.75266 & 0.75230 & 0.75266 & 0.75189\\ \hline
300000 & 0.75157 & 0.75225 & 0.75203 & 0.75183 & 0.75190 & 0.75191 & 0.75225 & 0.75157\\
300000 & 0.75188 & 0.75223 & 0.75239 & 0.75215 & 0.75201 & 0.75213 & 0.75239 & 0.75188\\
300000 & 0.75201 & 0.75264 & 0.75266 & 0.75209 & 0.75229 & {0.75234} & 0.75266 & \tblue{0.75201}\\
300000 & 0.75177 & 0.75227 & 0.75273 & 0.75194 & 0.75237 & 0.75221 & 0.75273 & 0.75177\\
300000 & 0.75256 & 0.75189 & 0.75215 & 0.75221 & 0.75165 & 0.75209 & 0.75256 & 0.75165\\ \hline
400000 & 0.75225 & 0.75225 & 0.75220 & 0.75194 & 0.75207 & 0.75214 & 0.75225 & 0.75194\\
400000 & 0.75207 & 0.75214 & 0.75206 & 0.75209 & 0.75199 & 0.75207 & 0.75214 & 0.75199\\
400000 & 0.75183 & 0.75218 & 0.75175 & 0.75226 & 0.75196 & 0.75200 & 0.75226 & 0.75175\\
400000 & 0.75209 & 0.75211 & 0.75209 & 0.75205 & 0.75226 & 0.75212 & 0.75226 & \tblue{0.75205}\\
400000 & 0.75218 & 0.75213 & 0.75206 & 0.75197 & 0.75256 & {0.75218} & 0.75256 & 0.75197\\ \hline
500000 & 0.75369 & 0.75356 & 0.75349 & 0.75358 & 0.75385 & 0.75363 & 0.75385 & 0.75349\\
500000 & 0.75354 & 0.75379 & 0.75365 & 0.75368 & 0.75359 & {0.75365} & 0.75379 & \tred{0.75354}\\
500000 & 0.75373 & 0.75384 & 0.75354 & 0.75342 & 0.75361 & 0.75363 & 0.75384 & 0.75342\\
500000 & 0.75368 & 0.75354 & 0.75314 & 0.75377 & 0.75360 & 0.75355 & 0.75377 & 0.75314\\
500000 & 0.75366 & 0.75354 & 0.75347 & 0.75366 & 0.75372 & 0.75361 & 0.75372 & 0.75347\\ \hline
600000 & 0.75333 & 0.75300 & 0.75322 & 0.75339 & 0.75305 & 0.75320 & 0.75317 & 0.75321\\
600000 & 0.75304 & 0.75324 & 0.75331 &0.753107 & 0.75276 & 0.75309 & 0.75310 & 0.75308\\
600000 & 0.75359 & 0.75331 & 0.75309 & 0.75310 & 0.75318 & 0.75325 & 0.75319 & 0.75316\\
600000 & 0.75335 & 0.75358 & 0.75326 & 0.75308 & 0.75332 & {0.75332} & 0.75331 & \tblue{0.75326}\\
600000 & 0.75334 & 0.75338 & 0.75309 & 0.75320 & 0.75325 & 0.75325 & 0.75323 & 0.75320\\
\hline
\end{tabular}
\caption{Experimental results of Algorithm~\ref{alg:1} for $d=9$.}
\end{table}

\begin{table}
\small
\begin{tabular}{|l|l|l|l|l|l|l|l|l|l|l||l|l|l|}
\hline
$n$&$r_0$&$r_1$&$r_2$&$r_3$&$r_4$&Avg&Max &Min\\
\hline
100000 & 0.77768 & 0.77860 & 0.77746 & 0.77780 & 0.77798 & \tblue{0.77790} & 0.77860 & 0.77746\\
100000 & 0.77726 & 0.77844 & 0.77850 & 0.77792 & 0.77734 & 0.77789 & 0.77850 & \tblue{0.77726}\\
100000 & 0.77782 & 0.77770 & 0.77760 & 0.77770 & 0.77740 & 0.77764 & 0.77782 & 0.77740\\
100000 & 0.77760 & 0.77734 & 0.77786 & 0.77732 & 0.77726 & 0.77748 & 0.77786 & 0.77726\\
100000 & 0.77734 & 0.77754 & 0.77826 & 0.77796 & 0.77724 & 0.77767 & 0.77826 & 0.77724\\ \hline
200000 & 0.77788 & 0.77814 & 0.77838 & 0.77824 & 0.77830 & 0.77819 & 0.77838 & 0.77788\\
200000 & 0.77804 & 0.77806 & 0.77841 & 0.77821 & 0.77799 & 0.77814 & 0.77841 & 0.77799\\
200000 & 0.77810 & 0.77819 & 0.77841 & 0.77794 & 0.77798 & 0.77812 & 0.77841 & 0.77794\\
200000 & 0.77838 & 0.77820 & 0.77800 & 0.77829 & 0.77869 & {0.77831} & 0.77869 & \tred{0.77800}\\
200000 & 0.77745 & 0.77820 & 0.77796 & 0.77827 & 0.77856 & 0.77809 & 0.77856 & 0.77745\\ \hline
300000 & 0.77805 & 0.77791 & 0.77801 & 0.77805 & 0.77807 & 0.77802 & 0.77807 & \tblue{0.77791}\\
300000 & 0.77801 & 0.77828 & 0.77797 & 0.77773 & 0.77789 & 0.77798 & 0.77828 & 0.77773\\
300000 & 0.77735 & 0.77793 & 0.77810 & 0.77805 & 0.77775 & 0.77784 & 0.77810 & 0.77735\\
300000 & 0.77817 & 0.77819 & 0.77777 & 0.77809 & 0.77817 & {0.77808} & 0.77819 & 0.77777\\
300000 & 0.77806 & 0.77789 & 0.77729 & 0.77816 & 0.77801 & 0.77788 & 0.77816 & 0.77729\\ \hline
400000 & 0.77782 & 0.77759 & 0.77809 & 0.77795 & 0.77780 & 0.77785 & 0.77809 & 0.77759\\
400000 & 0.77791 & 0.77826 & 0.77783 & 0.77759 & 0.77798 & {0.77791} & 0.77826 & 0.77759\\
400000 & 0.77789 & 0.77780 & 0.77757 & 0.77790 & 0.77793 & 0.77781 & 0.77793 & 0.77757\\
400000 & 0.77789 & 0.77777 & 0.77789 & 0.77801 & 0.77792 & 0.77789 & 0.77801 & \tblue{0.77777}\\
400000 & 0.77789 & 0.77746 & 0.77794 & 0.77772 & 0.77766 & 0.77773 & 0.77794 & 0.77746\\ \hline
500000 & 0.77770 & 0.77782 & 0.77748 & 0.77784 & 0.77754 & 0.77767 & 0.77784 & 0.77748\\
500000 & 0.77748 & 0.77744 & 0.77805 & 0.77779 & 0.77777 & 0.77771 & 0.77805 & 0.77744\\
500000 & 0.77812 & 0.77790 & 0.77765 & 0.77769 & 0.77754 & 0.77778 & 0.77812 & 0.77754\\
500000 & 0.77775 & 0.77765 & 0.77772 & 0.77772 & 0.77778 & 0.77772 & 0.77778 & 0.77765\\
500000 & 0.77816 & 0.77780 & 0.77766 & 0.77766 & 0.77790 & {0.77784} & 0.77816 & \tblue{0.77766}\\ \hline
600000 & 0.77793 & 0.77785 & 0.77769 & 0.77762 & 0.77780 & {0.77778} & 0.77775 & \tblue{0.77773}\\
600000 & 0.77772 & 0.77762 & 0.77764 & 0.77768 & 0.77786 & 0.77770 & 0.77770 & 0.77772\\
600000 & 0.77789 & 0.77788 & 0.77759 & 0.77758 & 0.77750 & 0.77769 & 0.77765 & 0.77760\\
600000 & 0.77748 & 0.77772 & 0.77745 & 0.77748 & 0.77754 & 0.77753 & 0.77755 & 0.77751\\

\hline
\end{tabular}
\caption{Experimental results of Algorithm~\ref{alg:1} for $d=10$.}
\end{table}

{\small
\begin{thebibliography}{99}

\bibitem{Alon1997} N. Alon. On the edge-expansion of graphs. {\em Combinatorics, Probability and Computing}, 6:145–152, 1997.
%


\bibitem{Bichot2011} C. E. Bichot, P. Siarry (Eds.). {\em Graph Partitioning}, Wiley, 2011.
%
\bibitem{bollobas1980} B. Bollob\'as, A probabilistic proof of an asymptotic formula for the number of labelled regular graphs, {\em Europ. J. Combinatorics}, 1 (1980) 311--316.

\bibitem{Bollobas1988} B. Bollob\'as, The isoperimetric number of random regular graphs, {\em European J. Combinatorics}. 9,  241–244, 1988.

\bibitem{Bollobas1982} B. Bollob\'as, W. Fern\'andez  de la Vega,  The diameter  of random regular graphs, {\em Combinatorica}, 2  125--134, 1982.
%
\bibitem{Brandes2009} Brandes, U., and, Fleischer, D.  Vertex bisection is hard, too. {\em Journal of Graph Algorithms and Applications}, 13(2), 119-131, 2009.
%
\bibitem{Bui1987} T. Bui, S. Chaudhuri, T. Leighton, and M. Sipser. 
Graph bisection algorithms
with good average case behavior. {\em Combinatorica}, 7:171–191, 1987.
%
\bibitem{Buser1984} P. Buser, On the bipartition of graphs. {\em Discrete Applied Mathematics}, 9  105-109, 1984.
%
\bibitem{Clarke1989} L. H. Clarke and R. C. Entringer. The bisection width of cubic graphs. {\em Bull. of the Australian Math. Society},  39(3), 389 - 396, 1989.



\bibitem{Delling2012} D. Delling, A.V. Goldberg, I. Razenshteyn and R.F. Werneck. 
{\em Exact combinatorial branch-and-bound for graph bisection}. In: Proc. of the 14th  ALENEX, pp. 30–44, 2012.
%
\bibitem{Diaz2010} J. D\'iaz and D. Mitsche, The cook-book approach to the differential equation method, {\em Computer Science Review}. 4 , 129--151, 2010.

\bibitem{Diaz2002} J. D\'iaz, J. Petit, and M. Serna. A survey of graph layout problems. {\em ACM Computing Surveys}, 34:313–356, 2002.
%
\bibitem{Diaz2003} J. Diaz, N. Do, M.J. Serna, N.C. Wormald. Bounds on the max and min bisection of random cubic and random 4-regular graphs, Theoretical Computer Science, Volume 307, Issue 3, 2003, Pages 531-547.
%
\bibitem{Diaz2007} J. D\'iaz, M. Serna, N.C. Wormald. Bounds on the bisection width for random d-regular graphs, {\em Theoretical Computer Science}, 382 (2),  120-130, 2007.
%
\bibitem{Diaz2017} J. Díaz and G. B. Mertzios. Minimum Bisection is NP-hard on Unit Disk Graphs. {\em Information and Computation}, 256: 83-92, 2017.
%
\bibitem{Fraire2014} H. Fraire, D. Ter\'an-Villanueva, N.C. ~García N.C., J.J.G.~Barbosa, 
E.R.~Angel and Y.G.~ Rojas.
Exact Methods for the Vertex Bisection Problem. In: O.~Castillo, P.~Melin, W.~Pedrycz and J.~Kacprzyk
(eds.) {\em Recent Advances on Hybrid Approaches for Designing Intelligent Systems}. Studies in Computational Intelligence, vol. 547. Springer, 2014. 
%
\bibitem{Feldmann11} A.E. Feldmann, P. Widmayer. {\em An $O(n^4)$ time algorithm to compute the bisection width of solid grid graphs}. In: Proc. of  19th ESA, 143-–154, 2011.
%
\bibitem{Gao2017} P. Gao, and N. Wormald. Uniform generation of random regular graphs. {\em SIAM Journal Computing}, 46, 1395--1427, 2017.
%
\bibitem{Garey1976} M. R. Garey, D. S. Johnson, and L. Stockmeyer. Some simplified NP-complete graph problems. {\em Theoretical Computer Science}, 1, 237--267, 1976.
%
\bibitem{Harper1964} L.~Harper. Optimal assignments of number to vertices. {\em Journal of SIAM}, 12, 131--135, 1964.

\bibitem{Harper1966} L.~Harper. Optimal numbering and isoperimetric problems.{\em Journal of Combinatorial Theory}, 1, 385--393, 1966
%
\bibitem{Jain2016} P. Jain, G. Saran, K. Srivastava. Branch and bound algorithms for vertex bisection minimization problems. In: {\em Advanced Computing and Communication Tech.}, 452, Springer, 2016.
%
\bibitem{Jansen2005} K. Jansen, M. Karpinski, A. Lingas, and E. Seidel. 
Polynomial time approximation schemes for max-bisection on planar and geometric graphs. 
{\em SIAM J. on Computing}. 35 (1)  110–119, 2005.
%
\bibitem{Janson2000} S. Janson, T. {\L}uczak and A. Ruci\'{n}ski, {\em Random Graphs}, John Wiley, 2000.
%
\bibitem{Karp1981} R.M. Karp and M. Sipser, {\em Maximum matchings in sparse random graphs}, Proc. 22  IEEE FOCS, 364–375, 1981.


\bibitem{Karypis1998} G. Karypis, V. Kumar, METIS Manual, Version 4.0, University of Minnesota, Department of Computer Science, 1998.
%
\bibitem{Kerningan70} B. Kernighan, S. Lin. An efficient heuristic procedure for partitioning graphs. {\em Bell Systems and  Technology Journal}, 49 (2),  291–307, 1970.
%
\bibitem{Khot2004} S. Khot. {\em Ruling out PTAS for graph min-bisection, densest subgraph and bipartite clique}. In Proc of 43 IEEE FOCS-04,  136–145, 2004.
%
%
\bibitem{Kolesnik2014} B. Kolesnik and N. Wormald. Lower Bounds for the Isoperimetric Numbers of Random Regular Graphs. {\em SIAM Journal on Discrete Math.}, 28(1), 553-575, 2014.
%
\bibitem{Kurtz1970} T.G. Kurtz. Solutions of ordinary differential equations as limits of pure Markov jump processes. {\em Journal of Applied Probability}, 7,  49--58, 1970.
%
\bibitem{Lichev2020} L. Lichev and D. Mitsche. On the minimum bisection of Random $3$-regular graphs.   {\em arXiv:2009.00598}, 2020.

\bibitem{Lyons2017} R. Lyons. Factors of IID on trees. {\em Combinatorics, Probability and  Computing}, 26(2):285–-300, 2017.
%
\bibitem{Mac1978} R. M. MacGregor. {\em On Partitioning a Graph: A Theoretical and Empirical Study}, PhD Thesis, University of California, Berkeley, 1978.

\bibitem{makover2010} Makover, E. and McGowan, Regular trees in random regular graphs, {\em arXiv:Math:0610858v2}, 2010.
%

\bibitem{Nakano2003} K. Nakano. Linear layouts of generalized hypercubes. {\em Intl. J. Foundations of Computer Science}, 14, 137–-156, 2003.
%
%
\bibitem{Racke2008} H. R\"{a}cke. {\em Optimal Hierarchical Decompositions for Congestion Minimization in Networks}. In: Proc. of the 40th ACM  STOC, 255--264, 2008. 
%
\bibitem{Sauerwald2011} T. Sauerwald,  and A. Stauffer.  {\em Rumor spreading and vertex expansion on regular graphs}. In ACM-SIAM SODA, 462--475. SIAM,  2011.
%
%
%
\bibitem{vanBevern2015} R. van Bevern, A.E. Feldmann, M. Sorge, and O. Such\'y.
On the Parameterized Complexity of Computing Balanced Partitions in Graphs.
{\em Theory of Computing Systems}, 57:1–-35,  2015.

\bibitem{Wormald1999} N.C. Wormald. The Differential Equation Method for random graph processes and
greedy algorithms. In: M. Karonski and H. PrXomel (Eds.), {\em Lectures on Approximation
and Randomized Algorithms}, PWN, Warsaw,  73--155, 1999.
%
\bibitem{Wormald1999b} N.C. Wormald. Models of random regular graphs. In: {Surveys in Combinatorics}, London Math. Soc. Lecture Note Ser. 267, CUP,  239--298, 1999.  
\end{thebibliography}}
\end{document}